\documentclass[11pt]{article}
\usepackage{amssymb}
\usepackage{epsfig}
\begin{document}
\title{{\bf{\Large Tolman temperature once again: A derivation from gravitational surface action}}}
\author{ 
 {\bf {\normalsize Bibhas Ranjan Majhi}$
$\thanks{E-mail: bibhas.majhi@iitg.ernet.in}}\\ 
{\normalsize Department of Physics, Indian Institute of Technology Guwahati,}
\\{\normalsize Guwahati 781039, Assam, India}
\\[0.3cm]
}
\maketitle

\begin{abstract}
The temperature distribution in presence of gravity, as measured by a local observer, is given by the Tolman expression. Here I derive the same only from the Gibbon's-Hawking-York surface term. In this process no {\it explicit} use of Einstein's equations of motion is done. Therefore, the present one is an {\it off}-shell analysis. Finally I discuss the importance and various implications of the derivation.
\end{abstract}
\vskip 9mm

\section{\label{Intro}Introduction}
   In $1930$, Tolman \cite{Tolman:1930zza,Tolman:1930ona} showed that the proper temperature of a system, composed of matter in thermodynamic equilibrium with a gravitational field, as measured by a local observer, satisfies the relation: $T\sqrt{-g_{tt}}$ = constant, where $T$ is the proper temperature and the other one is the square root of the $dt^2$ coefficient of the background metric. The last one is known as redshift factor. This implies that the temperature of a black hole horizon, measured by an observer which is very near to the horizon, diverges as $g_{tt}=0$ in this regime. It turns out that such an observation is consistent with the Unruh effect \cite{Unruh:1976db}. The temperature calculated from the Rindler frame, which is passing through a hyperbola very near to the light cone, is infinite as the acceleration of the fame for such path is very large. Moreover, for Hawking radiation \cite{Hawking:1974rv}, the constant on the right hand side of the relation is identified as the Hawking temperature as in the asymptotic infinity limit the red shift factor goes to unity. The importance of this result is that it helps us to relate the temperatures at each point of spacetime. Till then this remains as an integral part of gravitational theories.
   
   In the original derivations \cite{Tolman:1930zza,Tolman:1930ona}, the information about the external matter (radiation) plays an important role.  In one approach, the conservation of the energy-momentum tensor for the ideal fluid has been used while in others they came through {\it explicit} use of Einstein' s equations of motion. Also the Stefan-Boltzmann law for radiation was needed {\footnote{Very recently, in \cite{Gim:2015era} a correction to Toman temperature has been introduced for the ($1+1$) dimensional case in presence of trace anomaly in the enery-momentum tensor for external fields. In this analysis a modification to the usual Stefan-Boltzmann law has been found.}}. Moreover, a derivation, based on the extremum principle, was done by extremizing the entropy of the radiation.  In summary, such an analysis used both the information of gravity as well as matter and is an on-shell calculation in the sense that the Einstein's equations of motion were needed. 
   
   I already mentioned that the Unruh temperature obeys the Tolman relation and also note that the Rindler metric is not a solution of Einstein's equation. So the expression is not only restricted to a particular theory of gravity, rather it has much more applicability. Therefore one must expect an derivation of such relation which does not use the equations of motion. Moreover, for the Hawking effect the constant is just the Hawking temperature ($1/8\pi M$ for Schwarzschild black hole), which is purely gravitational contribution. It is the temperature of the event horizon as measured from infinity. In addition, the redshift factor is also a contribution from gravitational field. All these indicate that the Tolman relation might has a completely gravitational interpretation and if that is so, it is possible to find a independent derivation which deals only gravity; i.e. the external matter part should not play any role. 
The aim of the present paper is precisely in these directions.  Here I shall obtain the required relation purely from the surface term of the gravitational action. The action will be taken as the Gibbons-Hawking-York (GHY) surface term. 

   Motivation for considering only the surface term is as follows. It must be noted that the Einstein-Hilbert (EH) action, in a local frame, reduces to a total derivative term. Moreover, the equations of motion can be obtained purely from GHY term under certain prescription \cite{Padmanabhan:2004fq}. Most importantly, it has been observed that the calculation of this action on the horizon leads to horizon entropy of a black hole \cite{Paddy} (see also \cite{Padmanabhan:2012bs,Majhi:2013jpk}). This is not a new observation. Earlier, when Gibbons-Hawking  \cite{Gibbons:1976ue} obtained the entropy in the action principle, it must be noted that the contribution came from only GHY term as the EH part vanishes after imposition of the on-shell condition. Very recently, the same was obtained from the Noether charge corresponding to surface action \cite{Majhi:2012tf,Majhi:2012nq} and also a entropy function has been defined to find the entropy of an near horizon extremal black hole \cite{Majhi:2015pra}. All these indicate that the surface may contain all the information about the bulk term. Keeping in mind the earlier aim and the present motivation, in this paper I shall give an analysis to find Tolman temperature distribution by just considering the GHY action.  In this analysis no information about the external matter and the Einstein's equations of motion will be used. We shall see that such a derivation is much more simple and give more insight of the relation. 
   
   Let me now summarise the procedure to be adopted in this paper. In the original derivation the entropy of the radiation was extremized. Since, the horizon entropy is related to the GHY term, here I shall extremize it to achieve the goal. The background will be chosen as a static, spherically symmetric metric. The action will be first expanded in terms this metric coefficients and then the usual extemization procedure will be applied. This will lead to equations for metric coefficients. Finally, using them the Tolman expression will be derived. Here the Einstein's equations of motion will not be used explicitly. 
   
   The organization of the paper is as follows.  In section \ref{GHY}, a short introduction about the GHY action will be given. Next section is the main part of the paper which will contain a detailed analysis to reach to the Tolman form by extremizing the surface action. Finally in section \ref{Conclusions}, I shall conclude. An appendix, with details of the calculation, will also be added at the end.

\section{\label{GHY}GHY surface term: a brief introduction}
     Let us first introduce the GHY surface action. It must be noted that the EH action consists of both the first order and the second order derivatives of the metric $g_{ab}$ as the Ricci scalar is made of square of connection and its derivative. Therefore an arbitrary variation of the action leads to a boundary term containing both the variations of the metric as well as its derivative. So in order to obtain the equations of motion, using the least action principle, we need to fix both the metric and its derivative at the boundary. But it must be mentioned that use of both the boundary conditions is not a correct prescription. This has been avoided by adding a surface term to action in such a way that the total variation (i.e. EH plus the surface action) is free of total derivative of variation of the derivative of $g_{ab}$. Then one has to fix only the metric at the boundary. 
Although there is no unique choice of the boundary term \cite{Charap:1982kn}, people, in most of the time, choose the GHY surface term, introduced by Gibbons, Hawking and York. Such a term is foliation dependent; defined either on the timelike surface or on the spacelike surface {\footnote{An attempt has been made recently in \cite{Parattu:2015gga} to define a boundary term on an arbitrary null surface.}}. It has been observed that in this case one has to fix only the induced metric of the foliated surface at the boundary to obtain the Einstein's equations of motion \cite{Paddy}. 
     
     The GHY surface action is defined as
\begin{equation}
\mathcal{A}_{GHY} = -\frac{1}{8\pi G}\int_{\partial \mathcal{V}}\sqrt{|h_{(i)}|}d^3x \epsilon K_{(i)}~,
\label{2.01}
\end{equation}
where $\partial \mathcal{V}$ is the boundary surface of the full manifold $\mathcal{M}$. As stated earlier $\partial\mathcal{V}$ can be either timelike or spacelike, depending on the value of $\epsilon$. Here $h_{(i)}$ is the determinant of the induced metric of the surface $\partial\mathcal{V}$ and $K_{(i)}$ is the second fundamental. The expression of $K_{(i)}$, in terms of the unit normal $N^a_{(i)}$ to the boundary, is  
\begin{equation}
K_{(i)}=-\nabla_aN^a_{(i)}~.
\label{2.02}
\end{equation}
Remember that ``$i$'' in the subscript denotes the kind of surface (spacelike or timelike) we are choosing.  Also keep in mind that $\epsilon =+1$ refers to the timelike surface while $\epsilon=-1$ denotes a spacelike surface.
For our future purpose, we shall express the GHY action in the following form. Gauss's theorem helps us to write Eq. (\ref{2.01}) as
\begin{equation}
\mathcal{A}_{GHY} = -\int_{\mathcal{M}}d^4x\sqrt{-g}\mathcal{L}_{GHY} =-\frac{1}{8\pi G}\int_{\mathcal{M}}d^4x\sqrt{-g} \nabla_a\Big(K_{(i)}N^a_{(i)}\Big)~,
\label{2.03}
\end{equation}
where one has to consider $N^a_{(i)}N_{a(i)}=\epsilon$. I shall use the above form in the next section to achieve the main goal.

\section{\label{Tolman}Tolman temperature}
   The present section will show that it is possible to obtain the Toman temperature distribution from the surface part of the whole gravitational action. We first calculate the action under a static, spherically symmetric background and the we shall see that the extremization of it will lead to the required relation. For that let us consider a general spherically symmetric, static metric:
\begin{equation}
ds^2 = -e^{\nu}dt^2+e^{\mu}(dr^2+r^2d\theta^2+r^2\sin^2\theta d\phi^2)~,
\label{3.01}
\end{equation}
where $\mu$ and $\nu$ are functions of radial coordinate $r$ only. Below we shall now calculate the GHY action, given by the form (\ref{2.03}), for the above metric. 

    Note that for the present metric the relevant surfaces, which are denoted by subscript ``$i$'' in (\ref{2.03}), are $t=constant$ surface which is spacelike and $r=constant$, $\theta=constant$, $\phi=constant$ surfaces which are timelike.
Therefore, we first write the Lagrangian density in the following form:
\begin{eqnarray}
\sqrt{-g}\mathcal{L}_{GHY} &=& \frac{1}{8\pi G}\partial_a\Big(\sqrt{-g}K_{(i)}N^a_{(i)}\Big)
\nonumber
\\
&=& \frac{1}{8\pi G}\Big[\partial_a\Big(\sqrt{-g}K_{(t)}N^a_{(t)}\Big)+\partial_a\Big(\sqrt{-g}K_{(r)}N^a_{(r)}\Big)
\nonumber
\\
&+& \partial_a\Big(\sqrt{-g}K_{(\theta)}N^a_{(\theta)}\Big)+\partial_a\Big(\sqrt{-g}K_{(\phi)}N^a_{(\phi)}\Big)\Big]~.
\label{3.02}
\end{eqnarray}
Now for the metric (\ref{3.01}), the unit normal vectors to different surfaces turn out to be
\begin{eqnarray}
&&N^a_{(t)}=\Big(e^{-\nu/2},0,0,0\Big); \,\,\,\ N^a_{(r)} = \Big(0,e^{-\mu/2},0,0\Big); 
\nonumber
\\
&&N^a_{(\theta)} = \Big(0,0,\frac{e^{-\mu/2}}{r},0\Big); \,\,\,\ N^a_{(\phi)} = \Big(0,0,0,\frac{e^{-\mu/2}}{r\sin\theta}\Big)~.
\label{3.03}
\end{eqnarray}
Moreover, the metric coefficients are independent of $t$ and $\phi$ coordinates. Therefore, the first and last terms on the right hand side of (\ref{3.02}) will not contribute and it reduces to
\begin{equation}
\sqrt{-g}\mathcal{L}_{GHY} = \frac{1}{8\pi G}\Big[\partial_r\Big(\sqrt{-g}K_{(r)}N^r_{(r)}\Big) + \partial_\theta\Big(\sqrt{-g}K_{(\theta)}N^\theta_{(\theta)}\Big)\Big]~.
\label{3.04}
\end{equation}
Next using $K_{(i)}=-\nabla_aN^a_{(i)}=-(1/\sqrt{-g})\partial_a(\sqrt{-g}N^a_{(i)})$ we rewrite the above as
\begin{equation}
\sqrt{-g}\mathcal{L}_{GHY} = -\frac{1}{8\pi G}\Big[\partial_r\Big\{N^r_{(r)}\partial_r\Big(\sqrt{-g}N^r_{(r)}\Big)\Big\} + \partial_\theta\Big\{N^{\theta}_{(\theta)}\partial_{\theta}\Big(\sqrt{-g}N^\theta_{(\theta)}\Big)\Big\}\Big]~.
\label{3.05}
\end{equation}
Substituting (\ref{3.03}) and $\sqrt{-g}=(r^2\sin\theta)\exp[(3\mu+\nu)/2]$ in the above and after doing some simplifications we obtain
\begin{equation}
\sqrt{-g}\mathcal{L}_{GHY} = -\frac{\sin\theta}{8\pi G} e^{\frac{\mu+\nu}{2}}\Big[1+r\Big(3\mu'+2\nu'\Big)+r^2\Big(\mu''+\frac{\nu''}{2}\Big)+r^2\Big(\frac{\mu'^2}{2}+\frac{\nu'^2}{4}+\frac{3\mu'\nu'}{4}\Big)\Big]~,
\label{3.06}
\end{equation}
where the prime denotes the derivative with respect to radial coordinate.
Using this in (\ref{2.03}) and performing the integration in $\theta$ and $\phi$, one obtains the GHY action as
\begin{equation}
\mathcal{A}_{GHY}=\frac{1}{2G}\int dt dr  e^{\frac{\mu+\nu}{2}}\Big[1+r\Big(3\mu'+2\nu'\Big)+r^2\Big(\mu''+\frac{\nu''}{2}\Big)+r^2\Big(\frac{\mu'^2}{2}+\frac{\nu'^2}{4}+\frac{3\mu'\nu'}{4}\Big)\Big]~.
\label{3.07}
\end{equation}
Since the metric is static, the integration on time coordinate can also be done. In this case we shall choose the limits of integration by the following argument. It is well known that if the usual time coordinate is Euclideanised first and then it is assumed to be periodic, the zero temperature theory leads to the finite temperature theory where the inverse temperature of the system is the periodicity of the Euclidean time. Following the same path, let us assume that the time has periodicity $\beta_0=1/T_0$. Then the above will lead to
\begin{equation}
\mathcal{A}_{GHY}=\frac{1}{2G}\int dr  \frac{r^2e^{\frac{\mu+\nu}{2}}}{T_0}\Big[\frac{1}{r^2}+\frac{3\mu'+2\nu'}{r}+\mu''+\frac{\nu''}{2}+\frac{\mu'^2}{2}+\frac{\nu'^2}{4}+\frac{3\mu'\nu'}{4}\Big]~.
\label{3.08}
\end{equation}
In the below, I shall vary the above action for arbitrary variations of $\mu$ and $\nu$ to obtain equations of motion for these two variables.

      The variation of the GHY action with respect to $\mu$ and $\nu$ yields
\begin{eqnarray}
&&\delta\mathcal{A}_{GHY} = \frac{1}{2G}\int dr  \frac{r^2e^{\frac{\mu+\nu}{2}}}{T_0}\Big(\frac{\delta\mu+\delta\nu}{2}\Big)\Big[\frac{1}{r^2}+\frac{3\mu'+2\nu'}{r}+\mu''+\frac{\nu''}{2}+\frac{\mu'^2}{2}+\frac{\nu'^2}{4}+\frac{3\mu'\nu'}{4}\Big]
\nonumber
\\
&&+\frac{1}{2G}\int dr \frac{r^2e^{\frac{\mu+\nu}{2}}}{T_0}\Big[\frac{3\delta\mu'+2\delta\nu'}{r}+\delta\mu''+\frac{\delta\nu''}{2}+\mu'\delta\mu'+\frac{\nu'\delta\nu'}{2}+\frac{3\nu'\delta\mu'}{4}+\frac{3\mu'\delta\nu'}{4}\Big]
\label{3.09}
\end{eqnarray}
Ignoring the total derivative terms; i.e. imposing the conditions $\delta\mu=\delta\nu=0$ and $\delta\mu'=\delta\nu'=0$ at the boundary $r=constant$, in the above we find the following expression for variation of the surface action:
\begin{eqnarray}
\delta\mathcal{A}_{GHY} &=& \frac{1}{2G}\int dr \Big[\frac{d^2}{dr^2}\Big(\frac{r^2e^{\frac{\mu+\nu}{2}}}{T_0}\Big)-\frac{d}{dr}\Big\{\frac{r^2e^{\frac{\mu+\nu}{2}}}{T_0}\Big(\frac{3}{r}+\mu'+\frac{3\nu'}{4}\Big)\Big\}+\mathcal{B}\Big]\delta\mu
\nonumber
\\
&+& \frac{1}{2G}\int dr \Big[\frac{d^2}{dr^2}\Big(\frac{r^2e^{\frac{\mu+\nu}{2}}}{2T_0}\Big)-\frac{d}{dr}\Big\{\frac{r^2e^{\frac{\mu+\nu}{2}}}{T_0}\Big(\frac{2}{r}+\frac{\nu'}{2}+\frac{3\mu'}{4}\Big)\Big\}+\mathcal{B}\Big]\delta\nu~,
\label{3.10}
\end{eqnarray}
where $\mathcal{B}$ is given by
\begin{equation}
\mathcal{B} = \frac{r^2e^{\frac{\mu+\nu}{2}}}{2T_0}\Big(\frac{1}{r^2}+\frac{3\mu'+2\nu'}{r}+\mu''+\frac{\nu''}{2}+\frac{\mu'^2}{2}+\frac{\nu'^2}{4}+\frac{3\mu'\nu'}{4}\Big)~.
\label{3.11}
\end{equation}
For shake of completeness and clarity, a detailed derivation of Eq. (\ref{3.10}) has been presented in Appendix \ref{AppA}.
Therefore, setting the variation of the action equal to zero one obtains the equations of motion for $\mu$ and $\nu$ as
\begin{eqnarray}
&&\frac{d^2}{dr^2}\Big(\frac{r^2e^{\frac{\mu+\nu}{2}}}{T_0}\Big)-\frac{d}{dr}\Big\{\frac{r^2e^{\frac{\mu+\nu}{2}}}{T_0}\Big(\frac{3}{r}+\mu'+\frac{3\nu'}{4}\Big)\Big\}+\mathcal{B} = 0~;
\label{3.12}
\\
&&\frac{d^2}{dr^2}\Big(\frac{r^2e^{\frac{\mu+\nu}{2}}}{2T_0}\Big)-\frac{d}{dr}\Big\{\frac{r^2e^{\frac{\mu+\nu}{2}}}{T_0}\Big(\frac{2}{r}+\frac{\nu'}{2}+\frac{3\mu'}{4}\Big)\Big\}+\mathcal{B} = 0~.
\label{3.13}
\end{eqnarray}
Subtracting (\ref{3.13}) from (\ref{3.12}) we obtain
\begin{equation}
\frac{d^2}{dr^2}\Big(\frac{r^2e^{\frac{\mu+\nu}{2}}}{2T_0}\Big) - 2\frac{d}{dr}\Big(\frac{re^{\frac{\mu+\nu}{2}}}{T_0}\Big) - \frac{d}{dr}\Big(\frac{r^2e^{\frac{\mu+\nu}{2}}}{T_0}\frac{\mu'}{2}\Big) - \frac{d}{dr}\Big(\frac{r^2e^{\frac{\mu+\nu}{2}}}{T_0}\frac{\nu'}{2}\Big) = 0~.
\label{3.14}
\end{equation}
Now the inverse of the proper temperature of the system can be defined as
\begin{equation}
\frac{1}{T} = \int_0^{\beta_0} dt e^{\frac{\nu}{2}} = \frac{e^{\frac{\nu}{2}}}{T_0}~.
\label{3.15}
\end{equation}
Use of this and after some rearrangements, Eq. (\ref{3.14}) can be expressed in the following form:
\begin{equation}
\frac{d}{dr}\Big[r^2e^{\frac{\mu}{2}}\frac{d}{dr}\Big(\frac{1}{T}\Big)\Big] = \frac{d}{dr}\Big[\frac{r^2e^{\frac{\mu}{2}}}{T}\frac{\nu'}{2}\Big]~.
\label{3.16}
\end{equation}
Next I shall solve the differential equation under a boundary condition to reach my ultimate goal.

    Before doing so let me make some comments about it. In the original work of Tolman \cite{Tolman:1930zza}, the same equation was obtained (see, Eq. (41) and Eq. (51) of \cite{Tolman:1930zza}) by two ways. In one case he found this by extremizing the entropy of the radiation, whereas the same was also derived by extremizing the entropy of a perfect fluid.  In the second approach, the first law of thermodynamics was used. It also must be noted that for both the approaches, Tolman used Einstein's equations of motion as well as Stefan-Boltzmann law for radiation to achieve the above result. On the contrary, here I derived the relation by just extremizing the GHY surface action where no used of the above stated informations was needed. So, in this sense, my analysis is {\it off}-shell. 
    
   Now the solution of the differential equation (\ref{3.16}) turns out to be
\begin{equation}
\frac{d\ln T}{dr} = -\frac{1}{2}\frac{d\nu}{dr}+C_0\frac{Te^{-\frac{\mu}{2}}}{r^2}~,
\label{3.17}
\end{equation}
where $C_0$ is the constant of integration. To fix this constant impose the condition that the variation of the proper temperature; i.e. $dT/dr$ vanishes at $r=0$, although $T$ is not equal to zero. It yields $C_0=0$. Finally integrating once again we find
\begin{equation}
T=Ce^{-\frac{\nu}{2}}~,
\label{3.18}
\end{equation}
with $C$ is a constant. This is the well known Tolman relation for temperature distribution.
    
   As promised earlier, I showed that the Tolman temperature can be derived purely from the surface term of the action. Let me now discuss the significance and importance of the present analysis. As we know an accelerated observer also measures temperature of the Rindler horizon, known as Unruh effect.  Note that such a metric is not a solution of Einstein's equations. Furthermore, one can verify that the Unruh temperature obeys the Tolman relation. Therefore, it is quite apparent that Tolman relation is much more fundamental than the Einstein's equation and hence one should expect an {\it off}-shell derivation of such relation. This has precisely been shown here. In the whole analysis no {\it explicit} information of Einstein's equations of motion has been borrowed. Moreover, as I mentioned earlier, the surface action is related to the horizon entropy. So it is obvious that we are actually extremizing the gravitational entropy, instead of the radiation entropy. Hence the present analysis signifies that the Tolman relation is more than a temperature distribution of the external fields under gravitational background; rather it is actually a relation for gravity itself. It is consistent with the Hawking effect in which case the constant $C$ is identified as the Hawking temperature. This is the temperature of the event horizon as measured from infinity. But if an observer radially moves towards the horizon, the measured value is redshifted by the relation (\ref{3.18}). Note that in this case the observer dependent value depends only on the gravitational fields. That is all the entities, appearing in the relation, are the properties of gravity itself; no properties of external matter are playing any role.
   
   The last point I want to mention is that in the calculation the contribution from the GHY action for all the surfaces has been taken into account which led to the correct expression for Tolman temperature. The same has also been observed earlier in \cite{Padmanabhan:2002sha} and \cite{Majhi:2015pra} in the context of thermodynamics of horizon from the surface term.

\section{\label{Conclusions}Conclusions}   
       In this short paper, I showed that it is possible to find the well know Tolman temperature from the surface term of the gravitational action. Here, GHY has been considered. The difference between the original one and the present one is that we do not need any explicit information about the evaluation of spacetime; i.e. the Einstein's equation of motion as well as the existence of external matter. In this sense this analysis is an {\it off}-shell derivation. As I stated earlier, this has an important consequence in the context of Unruh effect. Moreover, since the GHY action is related to entropy of horizon and as the extremization of it led to the Tolman relation, one can argue that the expression is a consequence of the extremization of gravitational entropy. This completely a new observation as in earlier case \cite{Tolman:1930zza} the same was obtained by extremizing the radiation entropy.
       
       Finally, so far only the surface term has been taken into account. It would be interesting to look into the same problem by considering the EH action only and also in presence of both EH action and GHY term. In addition, let me mention that the present analysis is confined within the general theory of relativity (GR). But we know that the Tolman relation is beyond any particular theory; it is applicable to any theory of gravity. Therefore, one needs to clarify this point by considering the surface term beyond GR theory. Let me also point out that even the original like derivation for other theories of gravity is absent in literature. The investigations in these directions are going on.

\vskip 9mm
\noindent
{\bf{Acknowledgements}}\\
\noindent
The research of the author is supported by a START-UP RESEARCH GRANT (No. SG/PHY/P/BRM/01) from Indian Institute of Technology Guwahati, India.

\vskip 9mm
\section*{Appendix}
\appendix
\section{\label{AppA}Evaluation of Equation (\ref{3.10})}
\renewcommand{\theequation}{A.\arabic{equation}}
\setcounter{equation}{0} 
Note that the first term on the right hand side of Eq. (\ref{3.09}) has the structure which is like $(\dots)\times\delta\mu$ and $(\dots)\times\delta\nu$ with no terms containing the variations of derivatives of $\mu$ and $\nu$. So we do not need to do anything on this term. On the contrary, the second term contains the variation of the derivatives of our variables. Since in order to obtain the equation of motion from an action, one has to find the coefficients of only the variations of the variables, our aim will be to write the second term similar to the form: $(\dots)\times\delta\mu$, etc. In this process, like the usual least action principle, one generates total derivative terms. Therefore, let us now concentrate on the second term only. First rewrite this in the following form:
\begin{equation}
\frac{1}{2G}\int dr \frac{r^2e^{\frac{\mu+\nu}{2}}}{T_0}\Big[\Big(\frac{3}{r}+\mu'+\frac{3\nu'}{4}\Big)\delta\mu'+\Big(\frac{2}{r}+\frac{\nu'}{2}+\frac{3\mu'}{4}\Big)\delta\nu'+\delta\mu''+\frac{\delta\nu''}{2}\Big]~.
\label{App1}
\end{equation}
Now the first and second terms of the integrant can be expressed as
\begin{equation}
\frac{r^2e^{\frac{\mu+\nu}{2}}}{T_0}\Big(\frac{3}{r}+\mu'+\frac{3\nu'}{4}\Big)\delta\mu' = \frac{d}{dr}\Big[\frac{r^2e^{\frac{\mu+\nu}{2}}}{T_0}\Big(\frac{3}{r}+\mu'+\frac{3\nu'}{4}\Big)\delta\mu\Big] - \frac{d}{dr}\Big[\frac{r^2e^{\frac{\mu+\nu}{2}}}{T_0}\Big(\frac{3}{r}+\mu'+\frac{3\nu'}{4}\Big)\Big]\delta\mu~,
\label{App2}
\end{equation}
and 
\begin{equation}
\frac{r^2e^{\frac{\mu+\nu}{2}}}{T_0}\Big(\frac{2}{r}+\frac{\nu'}{2}+\frac{3\mu'}{4}\Big)\delta\nu' = \frac{d}{dr}\Big[\frac{r^2e^{\frac{\mu+\nu}{2}}}{T_0}\Big(\frac{2}{r}+\frac{\nu'}{2}+\frac{3\mu'}{4}\Big)\delta\nu\Big] - \frac{d}{dr}\Big[\frac{r^2e^{\frac{\mu+\nu}{2}}}{T_0}\Big(\frac{2}{r}+\frac{\nu'}{2}+\frac{3\mu'}{4}\Big)\Big]\delta\nu~,
\label{App3}
\end{equation}
respectively.
Similarly the other two terms turn out to be
\begin{equation}
\frac{r^2e^{\frac{\mu+\nu}{2}}}{T_0}\delta\mu'' = \frac{d}{dr}\Big[\frac{r^2e^{\frac{\mu+\nu}{2}}}{T_0}\delta\mu'\Big] - \frac{d}{dr}\Big[\delta\mu\frac{d}{dr}\Big(\frac{r^2e^{\frac{\mu+\nu}{2}}}{T_0}\Big)\Big] + \frac{d^2}{dr^2}\Big(\frac{r^2e^{\frac{\mu+\nu}{2}}}{T_0}\Big)\delta\mu~,
\label{App4}
\end{equation}
and
\begin{equation}
\frac{r^2e^{\frac{\mu+\nu}{2}}}{T_0}\frac{\delta\nu''}{2} = \frac{d}{dr}\Big[\frac{r^2e^{\frac{\mu+\nu}{2}}}{T_0}\frac{\delta\mu'}{2}\Big] - \frac{d}{dr}\Big[\frac{\delta\nu}{2}\frac{d}{dr}\Big(\frac{r^2e^{\frac{\mu+\nu}{2}}}{T_0}\Big)\Big] + \frac{d^2}{dr^2}\Big(\frac{r^2e^{\frac{\mu+\nu}{2}}}{T_0}\Big)\frac{\delta\nu}{2}~.
\label{App5}
\end{equation}
Substitution of (\ref{App2})--(\ref{App5}) in the second term of (\ref{3.09}); i.e. in (\ref{App1}) by ignoring the total derivative terms one obtains (\ref{3.10}).


\vskip 9mm
\begin{thebibliography}{99}

\bibitem{Tolman:1930zza} 
  R.~C.~Tolman,
  Phys.\ Rev.\  {\bf 35}, 904 (1930).
  
\bibitem{Tolman:1930ona} 
  R.~Tolman and P.~Ehrenfest,
  Phys.\ Rev.\  {\bf 36}, no. 12, 1791 (1930).

\bibitem{Unruh:1976db} 
  W.~G.~Unruh,
  Phys.\ Rev.\ D {\bf 14}, 870 (1976).
  
\bibitem{Hawking:1974rv} 
  S.~W.~Hawking,
  Nature {\bf 248}, 30 (1974).\\
  S.~W.~Hawking,
  Commun.\ Math.\ Phys.\  {\bf 43}, 199 (1975)
  [Commun.\ Math.\ Phys.\  {\bf 46}, 206 (1976)].

\bibitem{Gim:2015era} 
  Y.~Gim and W.~Kim,
  ``A Quantal Tolman Temperature,''
  arXiv:1508.00312 [gr-qc].
 
 \bibitem{Padmanabhan:2004fq} 
  T.~Padmanabhan,
  Braz.\ J.\ Phys.\  {\bf 35}, 362 (2005)
  [gr-qc/0412068].

\bibitem{Paddy}
 T.~Padmanabhan,
 ``Gravitation: Foundations and frontiers'', Chapter 15, [Cambridge, UK: Cambridge Univ. Pr. (2010)].
 
 \bibitem{Padmanabhan:2012bs} 
  T.~Padmanabhan,
  Gen.\ Rel.\ Grav.\  {\bf 44}, 2681 (2012)
  [arXiv:1205.5683 [gr-qc]].

\bibitem{Majhi:2013jpk} 
  B.~R.~Majhi and T.~Padmanabhan,
  Eur.\ Phys.\ J.\ C {\bf 73}, no. 12, 2651 (2013)
  [arXiv:1302.1206 [gr-qc]].
  
\bibitem{Gibbons:1976ue} 
  G.~W.~Gibbons and S.~W.~Hawking,
  Phys.\ Rev.\ D {\bf 15}, 2752 (1977).

\bibitem{Majhi:2012tf} 
  B.~R.~Majhi and T.~Padmanabhan,
  Phys.\ Rev.\ D {\bf 86}, 101501 (2012)
  [arXiv:1204.1422 [gr-qc]].
  
\bibitem{Majhi:2012nq} 
  B.~R.~Majhi,
  Adv.\ High Energy Phys.\  {\bf 2013}, 386342 (2013)
  [arXiv:1210.6736 [gr-qc]].

\bibitem{Majhi:2015pra} 
  B.~R.~Majhi,
  ``Entropy function from the gravitational surface action for an extremal near horizon black hole,''
  arXiv:1503.08973 [gr-qc].
  
\bibitem{Charap:1982kn} 
  J.~M.~Charap and J.~E.~Nelson,
  J.\ Phys.\ A {\bf 16}, 1661 (1983).
  
\bibitem{Parattu:2015gga} 
  K.~Parattu, S.~Chakraborty, B.~R.~Majhi and T.~Padmanabhan,
  ``Null Surfaces: Counter-term for the Action Principle and the Characterization of the Gravitational Degrees of Freedom,''
  arXiv:1501.01053 [gr-qc].
  
  
     
  
  






\bibitem{Padmanabhan:2002sha} 
  T.~Padmanabhan,
  Class.\ Quant.\ Grav.\  {\bf 19}, 5387 (2002)
  [gr-qc/0204019].
  
 

\end{thebibliography}
\end{document}